\documentclass[aps,prl,twocolumn,longbibliography,nopacs,floats,nofootinbib,superscriptaddress]{revtex4-1}
\usepackage{graphicx}
\usepackage{dcolumn}
\usepackage{bm}
\usepackage{amsmath}
\usepackage{color}
\usepackage{amssymb}
\usepackage{ulem}
\usepackage{xcolor}
\usepackage{subfigure}
\usepackage{multibib}

\definecolor{cardinal}{rgb}{0.6,0,0}
\definecolor{darkgreen}{rgb}{0,0.4,0}
\definecolor{golden}{rgb}{0.92, 0.7, 0}
\definecolor{midnight}{rgb}{0, 0, 0.5}
\definecolor{darkblue}{rgb}{0, 0, 0.7}

\def\he4{$^4$He}
\def\hel3{$^3$He}
\def\Am3{\AA$^{-3}$}
\def\beq{\begin{equation}}
\def\eeq{\end{equation}}

\newcommand{\be}{\begin{equation}}
\newcommand{\ee}{\end{equation}}
\newcommand{\bea}{\begin{eqnarray}}
\newcommand{\eea}{\end{eqnarray}}
\newcommand{\bse}{\begin{subequations}}
\newcommand{\ese}{\end{subequations}}

\begin{document}

\author{Egor Babaev}
\affiliation{Department of Physics, KTH Royal Institute of Technology, Stockholm SE-10691, Sweden}

\title{Topological order in higher composites}
\begin{abstract}
 { We introduce the concept of composite topological order in   multicomponent systems. In such a state topological order appears only in higher-than-usual composites, with no topological order in elementary fields. We propose that such a state can be realized in Bose-Fermi mixtures in ultracold atoms.  }

\end{abstract}

\maketitle
{ The goal of this work is to discuss topological order, nonetheless, let
  us first overview certain concepts in symmetry-breaking condensates.}
In condensed matter, the standard symmetry-breaking { condensates }appear as condensation of bosons or fermion pairs.
The recent extensive experiments that combined multiple probes  \cite{Grinenko2021state,shipulin2023calorimetric}
generated significant interest in electron quadruplets { condensation} and symmetry breaking appearing in ``higher-order" composite fields in the absence of symmetry breaking in individual fields. 
The experiments \cite{Grinenko2021state,shipulin2023calorimetric} reported the states principally different from superconductors. It is a system of multiple species of electrons, distinguished e.g. by band index $i,j$. The corresponding creation operator can be denoted as $c^\dagger_{\sigma i }$
where $\sigma$ is the spin or pseudospin index. In this state, there is no order in fields composed of a pair of electrons 
$<c_{\sigma i}c_{\alpha i}>$, however, there is a symmetry-breaking order in higher composites such as $<c_{\sigma i}c_{\alpha i}c_{\gamma j}^\dagger c_{\delta j}^\dagger>$.
The theoretical mechanism for that is 
the proliferation of composite topological defects without the proliferation of elementary topological defects in a system with a superconducting ground state (see e.g. \cite{babaev2004superconductor,babaev2002phase,Smiseth2005field,Kuklov2008deconfined,Bojesen2013time,Bojesen2014phase,Kuklov2006decoinfined,Kuklov2008deconfined,Svistunov2015}) .
Similarly, such fluctuations-based mechanisms can be used to stabilize charge-4e superconductivity with the order parameter $<c_{\sigma i}c_{\alpha i}c_{\sigma j}  c_{\alpha j} >$.
This was proposed to occur under certain conditions such as the melting of pair-density-wave orders \cite{agterberg2008dislocations,berg2009charge,Radzihovsky2009liquid} or in the presence of thermal fluctuations and other broken symmetries 
\cite{Herland2010phase}. 
{  Fluctuations-induced composite-order phases were also discussed in  \cite{Kuklov2006decoinfined,Kuklov2008deconfined,Motrunich2008,Chen2013deconfioned} in the context of the discussion of { superconductor-like} gauge theories associated with deconfinement transitions \cite{senthil2004deconfined}.
Discussions related to the suppression of fermion bilinear condensation appeared also in other contexts \cite{lu2023symmetric} (cf \cite{fidkowski2010effects}).}

Such { superfluid} composite orders can arise also at zero temperature via a quantum phase transition. The important mechanism for that is the proximity to multicomponent Mott insulating state shown in { the theoretical works }\cite{kuklov2003counterflow,kuklov2004commensurate,kuklov2004superfluid}.
In particular, as   numerically verified  in two-component Bose-Hubbard model \cite{soyler2009sign},
that there could be orders of the kind $<b_i b_j>\ne 0$ or $<b_i b_j^\dagger>\ne 0$ in the absence of condensation of individual Bose species  $<b_i>=<b_j>=0$.
The microscopic mechanism of the formation of $<b_i b_j^\dagger>$ order, in that case, is rooted in interaction effects on a lattice arresting total current, allowing only exchange of positions of particles of different kinds, leading to the appearance of counterflow order parameter $<b_i b_j^\dagger> \ne 0$ when
$<b_i>=<b_j>=0$. {  The first experiment realizing such arresting the individual flows of a bosonic mixture on a lattice, without arresting counter-flow was reported in \cite{zheng2024observation}.}

Note that in
a ``trivial" sense, order parameters
of many ``conventional super states of matter" are composites of more than two fermionic fields. I.e., atoms forming condensate of He${}^4$ are composites of two electronic and two protonic and neutronic fields or can be written in terms of even more quark fields. 
Hence, it is important to emphasize that
the composite order discussed here is { not } associated with the direct formation of four-fermion bound states. Hence, it is principally different also from  $\alpha$-particles condensaton \cite{ropke1998four} and from the generalization
of exciton condensates (direct bound states of electrons and holes).
Here, we are dealing with more subtle statistical correlations of particles that are not physically bound. 
An example that arises already in the two-component case is associated with how the system couples to the magnetic field. 
Due to the lack of direct binding of particles, the relative density variation causes the system's coupling to the magnetic field. The resulting effective model is similar to a Skyrme model \cite{Grinenko2021state,Garaud2022effective} in contrast to the direct coupling of vector potential to phase gradient in superconductors or to (at most) dipolar coupling in systems with direct binding of electrons and holes.
However, the difference becomes especially clear when there 
are more than two components. In three-component case with the order parameter  $<b_i b_j^\dagger> \ne 0$ when
$<b_i>=<b_j>=0$  \cite{babaev2023hydrodynamics,blomquist2021borromean}
  there are only two independent superfluid modes, but there are three phase difference variables and three types of vortices, i.e., superflows, are clearly not presentable as transport of locally bound particles of one component and holes in another component. 
An illustrative case arises in the presence of a thermal gradient. That is, 
consider N-component fermions labeled by index $i$ forming at low temperature
 a superconducting state with multiple gaps $|\Delta_i|e^{i\phi_i}$. 
Consider intercomponent Josephson coupling breaking symmetry down to $U(1)\times Z_2$ \cite{Carlstrom2011length,Grinenko2020superconductivity}.
 The additional $Z_2$ symmetry breaking is caused by frustrated Josephson coupling terms $|\Delta_i||\Delta_j|\cos (\phi_i-\phi_j)$  that produce two energetically equivalent ground state phase-difference locking  $(\phi_i-\phi_j)\ne 0, \pi$. This arises when the prefactors of Josephson terms are positive.
Under certain conditions, at elevated temperatures, the system undergoes a phase transition into fermion quadrupling condensate characterized by the order only in the phase differences but not in individual phases, described by the order parameter $<c_{\sigma i}c_{\alpha i}c_{\sigma j}^\dagger c_{\alpha j}^\dagger>$ \cite{Bojesen2013time,Bojesen2014phase,Grinenko2021state,shipulin2023calorimetric}. Because of intercomponent Josephson coupling
in that state, there is no DC dissipationless charge transfer. In a conventional sense, it is dissipative.
However, it has other types of ``dissipationless" phenomena.
First, inhomogeneities   create macroscopic
persistent local currents \cite{Garaud2022effective}
observation of which in a state with dissipative DC current was reported in \cite{Grinenko2021state}.
Another effect we can point to is the dissipationless current arising as a reaction to a thermal gradient. Namely, the  inter-component phase difference $(\phi_i-\phi_j)$ in this state is in general temperature-dependent \cite{garaud2016thermoelectric,silaev2015unconventional,Carlstrom2011length}.
Hence, in response to a stationary thermal gradient, there will be a stationary gradient of the phase differences $\nabla(\phi_i-\phi_j)$. This means that there is a persistent counter-flow of components.
Consider the counterflow in x-direction, defined as some components flowing in positive $\hat{x}$ direction and some counterflowing in the opposite $-\hat{x}$ direction. Adding a magnetic field in $-\hat{z}$ direction results in opposite drifts of counterflowing components:  drifting in $ \hat{y}$
and  drifting in $-\hat{y}$.
Hence, that can be viewed as a ``color" Hall-like effect with carriers in different bands drifting in opposite transverse to the thermal gradient directions.

{  The above review highlights that}, there are many interesting properties in condensates where there is no algebraic order and no broken symmetry\footnote{Note that strictly speaking, the so-called breaking of a local gauge symmetry (i.e. in a superconductor) is not truly symmetry breaking
\cite{elitzur1975impossibility}, and even the ordinary superconductor can be described in terms of topological order, see e.g. the recent discussion in  
\cite{Svistunov2015,babaev2023hydrodynamics} }  arising at the level of the simplest possible fields, made of bosons or pairs of fermions but there is order in higher composites, such as four-electron fields.  
{  These findings motivate posing the question} are the possible closest counterparts of that for topological systems? Which orders are possible in multicomponent systems where ``elementary fields" do not form topological order, but in higher composite fields do (without forming ``direct" bound states such as excitons \cite{eisenstein2014exciton}, or topological order in certain spin systems \cite{fradkin2013field,nietner2017composite}).

 As  the simplest case, consider, for example, N-component mixture of Bosons and Fermions, forming ``Borromean" countercurrent order: i.e., where co-directed flows are arrested, but counter-flows are possible; for a superfluid counterpart of this order, see \cite{babaev2023hydrodynamics}. Then, a Quantum Hall effect is possible for the fermionic mode corresponding to the counter-propagation of electrons and bosons 
\footnote{ 
This has further generalizations { to more complicated fermionic composite fields. This would be a counterpart of complicated bosonic composite orders that
were demonstrated earlier}
, such as 
$<b_i b_j^\dagger b_j^\dagger>$  \cite{Dahl2008preemptive} or higher-composite superconducting order 
$<c_{\sigma i}c_{\alpha i}c_{\sigma j}^\dagger c_{\alpha j}^\dagger c_{\sigma j}^\dagger c_{\alpha j}^\dagger>$
discussed in \cite{Herland2010phase}.}.

\begin{figure}
    \centering
\includegraphics[width=0.3\textwidth]{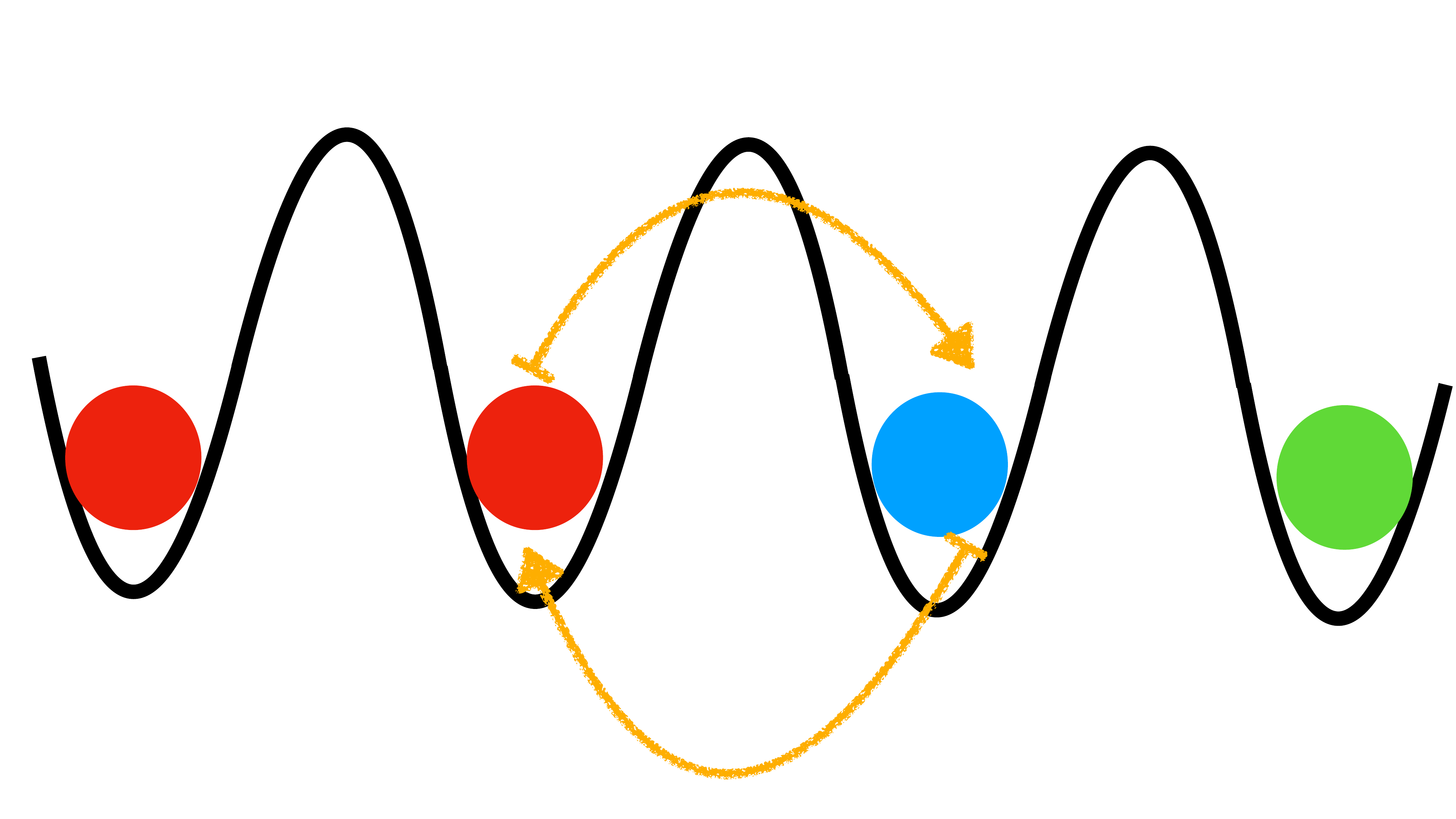}
    \caption{ Schematic depiction of a state with multiple species of particles denoted by different colors, residing in a lattice potential and having different inter- and inter-partice interactions. The state is a Mott insulator for total current but still permits the exchanges of particle's positions:  i.e. relative motion degrees of freedom are not arrested by interparticle interactions. { Here were are primarily interested in counterpropogating bosons and fermions,} for superfluid examples of bosonic counterflows near Mott transition see \cite{kuklov2003counterflow,kuklov2004commensurate,kuklov2004superfluid,soyler2009sign,Sellin18superfluiddrag,blomquist2021borromean}.}
    \label{HHM}
\end{figure}

{
For another example, consider a generalization of the Su-Schrieffer-Heeger (SSH) model for the case of an N-component mixture of bosons and fermions in one-dimensional elastic lattice with spacing $a$.

To generalize the SSH model, we start with an interacting multicomponent Hamiltonian
\begin{eqnarray}
    H_0&=
&-t\sum_{<{  ij}> l \sigma} (a^\dagger_{{  i}l \sigma}a_{{ j} l \sigma} +H.c.)\\ 
&+&\frac{1}{2}
\sum_{{  ij} \sigma \sigma ' l k} V_{\sigma \sigma ' l k}({ r}_{  i}-{  r}_{  j})n_{{  i}l\sigma}n_{{  j}k\sigma '}
\end{eqnarray}
Here $l,\sigma$ are the component and soon indices $n_{ l i}=a^\dagger_{{ l i}} a_{l i}$
The inter- and intra-component interaction can be chosen to make the system is a Mott insulator with respect to total currents, still permitting relative currents \cite{kuklov2003counterflow,kuklov2004commensurate}.
Here, we are interested in a   topological counterpart of N-component Borromean systems \cite{blomquist2021borromean,babaev2023hydrodynamics}. We consider the case where the underlying lattice arrests the total flow (see Fig. \ref{HHM}), that motivates an effective model that has only the relative motion (compare with the discussion in superfluid context \cite{babaev2023hydrodynamics}, where the arrest of total current can be described in terms of special gauge invariance), to that end we write\footnote{Formulation of this model relies on the numerical results that in three-component quantum system one can arrest total current without arresting relative current \cite{blomquist2021borromean}. More general models are possible with more complicated non-arrested counterflows such as $(Mq_la-Kq_ka)$, where $M,K$ are integers.}

\begin{eqnarray}
    H=\sum_{l
    \ne k} \epsilon_{lk} (q_la-q_ka)\sigma_x
\end{eqnarray}

When the lattice is ``elastic," the system has instability similar to the dimerization of the original SSH model.
Then one has hopping matrix elements become alternating
$t_{1,2}=t\pm \delta t$.
In this case a gap opens for the relative-motion modes, yielding the spectrum
\begin{eqnarray}
\epsilon_{lk\pm}=\pm\sqrt{ v^2(p_l-p_k)^2 +m^2}
\end{eqnarray}
and the system is topological with respect to composite modes, and has topologically nontrivial soliton states.
}

For a different type of example, we propose a  generalization of the Haldane-Hubbard model for several species of particles (that can include both fermions and bosons) labeled by the index $l$ (for the phase diagram of single-component Haldane-Hubbard model, see \cite{imrivska2016first,vanhala2016topological,tupitsyn2019phase})

\begin{eqnarray}
    H_0&=
&-t_1\sum_{<{\bf ij}> l \sigma} (a^\dagger_{{\bf i}l \sigma}d_{{\bf j} l \sigma} +H.c.)\\ 
  &-& t_2
e^{i\eta_{ij}\phi}\sum_{<<{\bf ij}>> l \sigma} (a^\dagger_{{\bf i}l \sigma}a_{{\bf j} l \sigma}+
d^\dagger_{{\bf i}l \sigma}d_{{\bf j} l \sigma} +H.c.)\\
&+&
D_l \sum_{{\bf i} \sigma} \xi ({\bf i}) n_{{\bf i}l\sigma} -
\sum_{{\bf i} \sigma} \mu_{l \sigma}n_{{\bf i}l\sigma} \\
&+&\frac{1}{2}
\sum_{{\bf ij} \sigma \sigma ' l k} V_{\sigma \sigma ' l k}({\bf r}_{\bf i}-{\bf r}_{\bf j})n_{{\bf i}l\sigma}n_{{\bf j}k\sigma '}
\end{eqnarray}
Here the two sublattices are labeled by A and B $\xi (i \in A)=+1, \xi (i \in B)=-1$ (see Fig. (\ref{HHM}) $\mu_{l \sigma}$ is chemical potential for a given spin component.
The terms with prefactors $t_1$ and $t_2$ correspond to the nearest- and next-to-nearest-neighbor hopping.
The last term corresponds to inter- and intra-species interactions.

\begin{figure}
    \centering
    \includegraphics[width=0.5\textwidth]{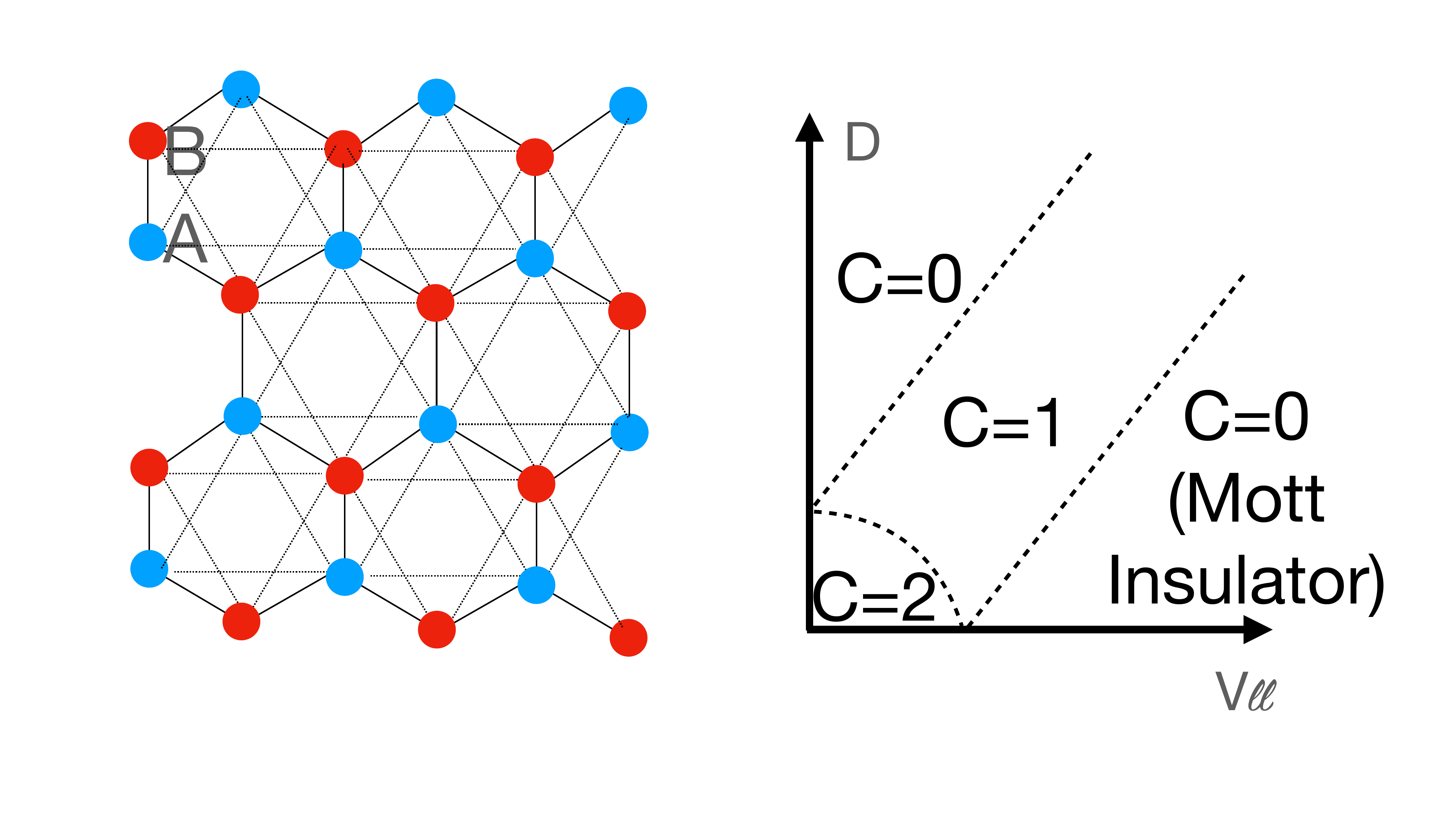}
    \caption{Schematic depiction of the two sub-lattice and phase diagram of single-component Haldane-Hubbard model \cite{imrivska2016first,vanhala2016topological,tupitsyn2019phase}.}
    \label{cartoon1}
\end{figure}

In a single-component Haldane Hubbard a significantly large interspecies interaction drives the system to a non-topological Mott insulator state  as schematically shown on Fig. (\ref{cartoon1}) \cite{imrivska2016first,vanhala2016topological,tupitsyn2019phase}. 

Again, our primary interest is Borromean systems \cite{blomquist2021borromean,babaev2023hydrodynamics}. When there is no intercomponent interaction $V_{\sigma \sigma ' l k} =0, ({k\ne l})$ the model represents N replicas of decoupled Haldane-Hubbard models.
However   new phases  arise with nonzero
intercomponent coupling $V_{\sigma \sigma ' l {k\ne l}} \ne 0$.
As demonstrated numerically in \cite{blomquist2021borromean}, by tuning the inter-  and intra-component interaction, one can make a system with more than two components a Mott insulator with respect to the total current but still retain the relative motion of different particles 
as schematically shown on Fig (\ref{HHM}). In the generalized model that we propose, this results in a topological state in terms of the relative motion of $l$ and $k\ne l$ components. This constitutes a topological counterpart of the composite superfluid order.

Importantly,   { this results in a particular type of fractionalization. That is, consider }  e.g.  a multicomponent boson-fermion mixture. { Then,} a fermion can take part in a normal or superfluid counter-flow with different species of fermion and also be a part of a topological boundary mode, corresponding to topological counterflow associated with exchanging positions with bosons. { Likewise for a counterfluid mixture of two species of bosons and fermions, a boson can participate in a counterflow superfluidity in the bulk and in a topological counterflowing mode with fermion on a boundary. In a coarse-grained sense, the boundary mode is composed of counterflowing fermions and two fractions of different bosons. }

Finally, we note that component indexes can 
have different origins. The most straightforward case is a mixture of different species of  fermions or bosons. The realization of the ordinary topological systems in ultracold atoms has attracted attention  \cite{jotzu2014experimental},
while there is great control over Mott transitions and preparation of various mixtures \cite{greiner2002quantum,bloch2008many,bloch2012quantum}. 
{ The recent work reported realizing the counter-flow-only mixture of ultracold atoms \cite{zheng2024observation}, which opens up the possibility of realizing the states we discuss in this work by creating the proposed optical lattices.} Similarly, it can represent other quantum numbers e.g. originating from multiple electronic bands.
 
\section{Acknowledgements}
The author thanks Johan Carlstrom, Nikolay Prokofiev, T.H. Hansson, Emil Bergholtz  and especially Boris Svistunov for many useful discussions.
 This work is supported by the Swedish Research Council Grants 2016-06122 and 2022-04763, by  Olle Engkvists Stiftelse and by the Knut and Alice
Wallenberg Foundation through the Wallenberg Center
for Quantum Technology (WACQT). 
%

\end{document}